\documentclass[preprint,review,12pt]{elsarticle}
\pdfoutput=1

\usepackage{graphicx}
\usepackage{amssymb}
\usepackage{amsthm}
\usepackage[outdir=./]{epstopdf}
\usepackage{lineno}
\usepackage{mathtools}
\usepackage{color, soul}
\usepackage{amsthm}

\newtheorem*{definition*}{Definition}

\journal{Sensors and Actuators Reports}

\begin{document}

\begin{frontmatter}

\title{Dual oxygen and temperature luminescence learning sensor with parallel inference}

\author[zhaw,toelt]{Francesca Venturini\corref{cor1}}
\author[toelt]{Umberto Michelucci}
\author[zhaw]{Michael Baumgartner}

\cortext[cor1]{Corresponding author. E-mail: francesca.venturini@zhaw.ch}

\address[zhaw]{Institute of Applied Mathematics and Physics, Zurich University of Applied Sciences, Technikumstrasse 9, 8401 Winterthur, Switzerland}
\address[toelt]{TOELT llc, Birchlenstr. 25, 8600 D\"ubendorf}

%\begin{frontmatter}
\begin{abstract}
A well-known approach to the optical measure of oxygen is based on the quenching of luminescence by molecular oxygen. The main challenge for this measuring method is the development of an accurate mathematical model. Typically, this is overcome by using an approximate empirical model where these effects are parametrized ad hoc. The complexity increases further if multiple parameters (like oxygen concentration and temperature) need to be extracted, particularly if they are cross interfering. The common solution is to measure the different parameters separately, for example, with different sensors, and correct for the cross interferences.
In this work, we propose a new approach based on a learning sensor with parallel inference. We show how it is possible to extract multiple parameters from a single set of optical measurements without the need for any {\sl a priori} mathematical model, and with unprecedented accuracy. We also propose a new metrics to characterize the performance of neural network based sensors, the Error Limited Accuracy. The proposed approach is not limited to oxygen and temperature sensing. It can be applied to the sensing with multiple luminophores, whenever the underlying mathematical model is not known or too complex.
\end{abstract}


\end{frontmatter}

\section{Introduction}
\label{Introduction}

The simultaneous determination of multiple physical quantities can be very advantageous in many sensor applications, for example, when an in-situ or a remote acquisition is required. 
If the physical effect on which the measurement method is based presents cross-interference of more than one quantity, their simultaneous determination becomes a necessity.
Optical luminescence sensing is particularly attractive for multiple sensing. Using the same measuring principle, several optical elements, like optical fibers and detectors, can be shared in the setup for the detection of more than one parameter, thus allowing a compact and simple sensor design.

The typical approaches to multiple sensing are based on either the use of a single luminescence indicator (luminophore), whose luminescence is sensitive to more than one physical quantity, or the use of several luminophores, one for each quantity, embedded in a substrate and placed in close physical proximity \cite{Stich2010,Borisov2011novel,Kameya2014,Wang2014,Santoro2016,Biring2019}. To be able to determine each quantity separately, it may be necessary to determine more than one optical property (e.g., absorption spectrum, emission spectrum, luminescence intensity, decay time). Another possibility is to measure one single optical property using special detection schemes that take advantage of the emission properties of the used luminophores \cite{Wang2014,Biring2019,Collier2013,Stehning2004,Jorge2008,Moore2006}. 

The problem of dual sensing is particularly relevant in applications that involve oxygen sensing. The determination of oxygen partial pressure is of great interest in numerous fields, like medicine, biotechnology, environmental monitoring, or chemistry since oxygen plays an important role in many processes \cite{Papkovsky2013,Wang2014}. One of the most used optical measuring approaches uses the effect of the dynamical luminescence quenching by oxygen molecules. The measuring principle is based on the measurement of the luminescence of a specific luminophore, whose intensity and decay time are reduced due to collisions with molecular oxygen \cite{Lakowicz2006}.

Sensors based on this principle must rely on approximated empirical models to parametrize the dependence of the measured sensing quantity (e.g., luminescence intensity or decay time) on influencing factors. Among these, the temperature is the factor with the strongest influence since both the luminescence and the quenching phenomena are strongly temperature-dependent. Therefore, in any optical oxygen sensor, the temperature must be continuously monitored, most frequently with a separate sensor, and used to correct the calculated oxygen concentration \cite{Li2015}. This task can be difficult in practical implementation and may become a significant source of error in sensors based on luminescence sensing. Another disadvantage of this approach is that the parametrization of the sensor response with temperature is system-specific since it depends  on how the sensing element was fabricated and on the sensor itself \cite{Xu1994,Draxler1995,Hartmann1996,Mills1998,Badocco2008,Dini2011}.

In this work, we propose a revolutionary approach based on neural networks for parallel inference. The method enables accurate dual-sensing, using one single luminophore, and measuring a single quantity.
Instead of describing the response of the sensor as a function of the relevant parameters through an analytical model, a neural network  was designed and trained to predict both oxygen concentration and temperature simultaneously.
This new approach is based on multi-task learning (MTL) neural network architectures. These are characterized by common hidden layers, whose output is then the input of multiple branches of task-specific hidden layers. MTL architectures were chosen because they can learn correlated tasks \cite{Argyriou2006, Thrun1996, Caruana1997, Zhang2017, Baxter2000, Thung2018}. In a previous purely theoretical study that used only synthetic data, the authors showed that MTL architectures can be flexible enough to address multi-dimensional regressions problems \cite{Michelucci2019_2}. This work demonstrates for the first time that this is indeed true by building and characterizing a real physical optical sensor based on this principle.

To train the MTL neural network and to test the performance of the sensor on unseen data a very large amount of data is needed. Since the collection  cannot be performed by hand a fully automated data collection setup was developed and used to both vary the sensor environment conditions (gas concentration and temperature) and to collect the sensor response. 

This work proposes a paradigm shift from the classical description of the response of a sensor through an approximate model to the use of MTL sensor learning thanks to neural networks. 
These will learn the complex inter-parameter dependencies and sensor-specific response characteristics from a large amount of data automatically collected. This new method will enable to build sensors even if the response of the system to the physical quantities is too complex to be comfortably described by a mathematical model.

\section{Methods}
\label{sec:methods}

\subsection{Luminescence Quenching for Oxygen Determination}
\label{Theory}

Luminescence-based oxygen sensors usually consist of a luminophore whose luminescence intensity and decay time decrease for increasing O$_2$ concentrations. This reduction is due to collisions of the excited luminophore with molecular oxygen, which thus provides a radiationless deactivation process (collisional quenching). 
In the case of homogeneous media characterized by an intensity decay which is a single exponential, the decrease in intensity and lifetime are both described by the Stern-Volmer (SV) equation \cite{Lakowicz2006}
\begin{equation}
\frac{I_0}{I}=\frac{\tau_0}{\tau}=1+K_{SV} \cdot \left[O_2\right]
\label{SVe}
\end{equation}
where $I_0$ and $I$, respectively, are the luminescence intensities in the absence and presence of oxygen, $\tau_0$ and $\tau$ the decay times in the absence and presence of oxygen, $K_{SV}$ the Stern–Volmer constant and $\left[O_2\right]$ indicates the oxygen concentration.

For practical applications, the luminophore needs to be embedded in a supporting substrate, frequently a polymer. As a result, the SV curve deviates from the linear behavior of Eq. (\ref{SVe}). This deviation can be due, for example, to heterogeneities of the micro-environment of the luminophore, or to the presence of static quenching \cite{Wang2014}. A proposed scenario describes this non-linear behavior as due to the presence in the substrate of two or more environments, in which the luminescence is quenched at different rates \cite{Carraway1991,Demas1995}. This multi-site model describes the SV curve as the sum of $n$ contributions as
\begin{equation}
\frac{I_0}{I}=\bigg[ \sum_{i=1}^n
\frac{f_i}{1+K_{SVi} \cdot \left[O_2\right]}
\bigg]^{-1}
\label{SVe2}
\end{equation}
where $f_i$'s are the fractions of the total emission for each component under unquenched conditions, and $K_{SVi}$'s are the associated effective Stern–Volmer constants. Depending on the luminophore and on the substrate material, the models proposed in the literature may be even more complex \cite{Demas1995,Hartmann1995,Mills1999}.

In most industrial and commercial sensors, the decay time $\tau$ is frequently preferred to intensity measurement because of its higher reliability and robustness \cite{Wei2019}. The determination of the decay time is done most easily in the frequency domain by modulating the intensity of the excitation.  As a result, the emitted luminescence is also modulated but shows a phase shift $\theta$ due to the finite lifetime of the excited state. This method has the additional advantage of allowing very simple and low-cost implementation.

Although the multi-site model was introduced for luminescence intensities, it is frequently also used to describe the oxygen dependence of the decay times \cite{Demas1995,Quaranta2012}. Therefore, in the simplest case of a two-sites scenario, the model can be rewritten in terms of phase shift as \cite{Michelucci2019}
\begin{equation}
\begin{aligned}
\frac{\tan \theta_0 (\omega, T)}{\tan \theta (\omega, T, [O_2])}=
 & \bigg( \frac{f (\omega , T) }{1+K_{SV1} (\omega , T) \cdot \left[O_2\right]}+ \\
&\frac{1-f (\omega , T) }{1+K_{SV2} (\omega , T) \cdot \left[O_2\right]} \bigg)^{-1} \\
\label{theta_full}
\end{aligned}
\end{equation}
where $\theta_0$ and $\theta$, respectively, are the phase shifts in the absence and presence of oxygen, $f$ and $1-f$ are the fractions of the total emission for each component under unquenched conditions, $K_{SV1}$ and $K_{SV2}$ are the associated Stern–Volmer constants for each component, and $\omega$ is the angular modulation frequency. It is to be noted that the quantities $\theta_0$, $f$, $K_{SV1}$, and $K_{SV2}$ are all non-linearly temperature dependent \cite{Ogurtsov2006,lo2008,Zaitsev2016}. Additionally, if the modulation frequency is varied, they may show a frequency dependence, an artifact due to the approximate nature of the model. Finally, Eq. (\ref{theta_full}) needs to be inverted to determine $[O_2]$ from the measured quantity $\theta$.

The proposed approach not only solves the difficulties of finding an approximate mathematical model for a complex system, but also allows the determination of multiple quantities simultaneously. Even if it is an approximate description, however, the structure of Eq. (\ref{theta_full}), remains relevant to understand the structure of the data and optimize the architecture of the neural network.

\subsection{Experimental Procedure}
\label{Experimental}

\begin{figure}[t!]
\centering
\includegraphics[keepaspectratio, width=8.3cm]{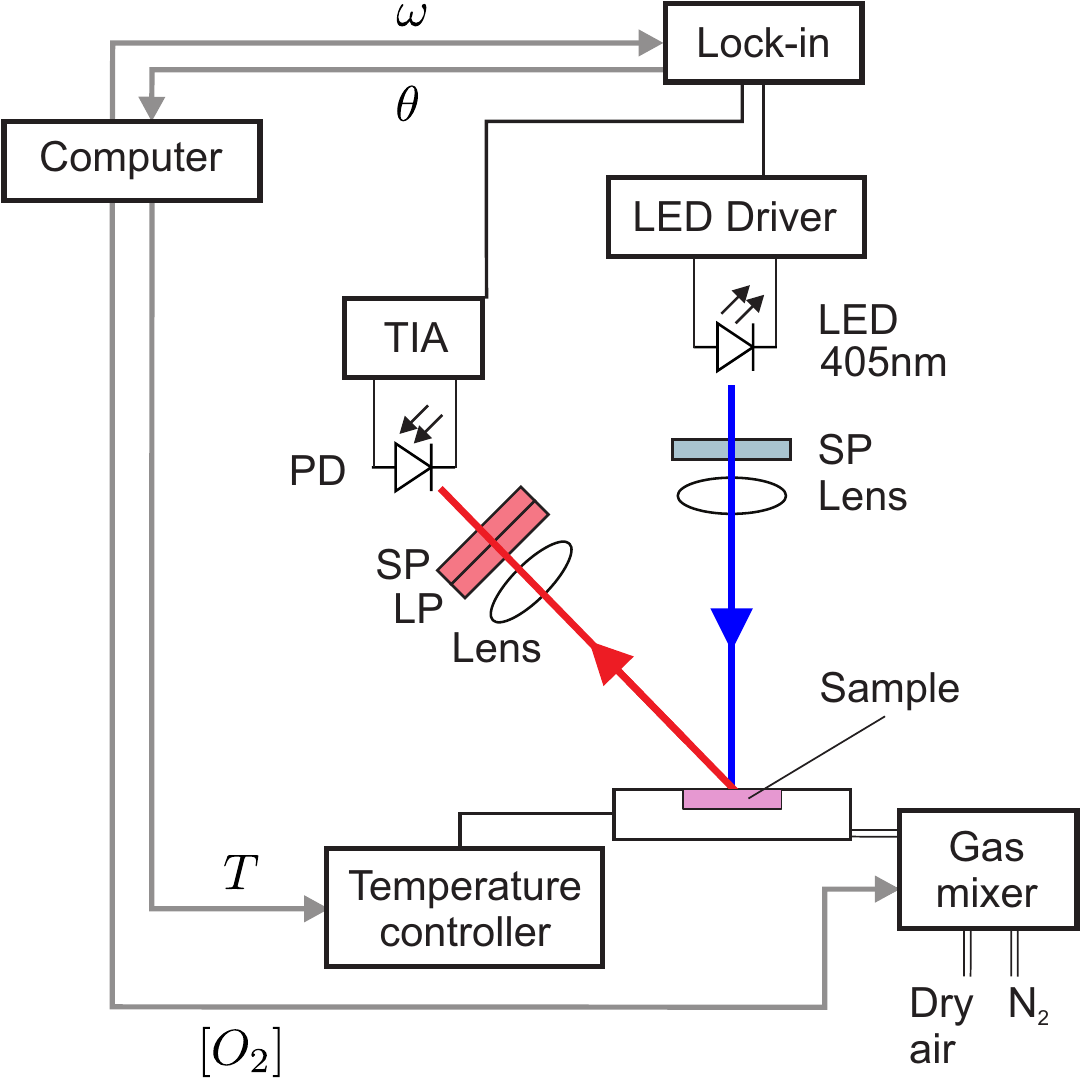}
\caption{Schematic diagram of the experimental setup. Blue indicates the excitation optical path, red the luminescence one. SP: shortpass filter; LP: longpass filter PD: photodiode; TIA: trans-impedance amplifier.}
\label{fig:setup}
\end{figure}

The optical setup used in this work for the luminescence measurements is shown schematically in Fig. \ref{fig:setup}. To be able to acquire a large number of data, the program for both the instrument control and the data acquisition was written using the software LabVIEW by National Instruments. The acquisition procedure is described in detail in Section \ref{Data}.

\subsubsection{Experimental Setup}

The sample used for the characterization and test is a commercially available Pt-TFPP-based oxygen sensor spot (PSt3, PreSens Precision Sensing).
To control its temperature, the sample was placed in good thermal contact with a copper plate, set in a thermally insulated chamber. The temperature of this plate was adjusted and stabilized using a Peltier element with a temperature controller (PTC10, Stanford Research Systems). The thermally insulated chamber was connected to a self-made gas-mixing apparatus, which enabled to vary the oxygen concentration between 0 $\%$ and 20 $\%$ vol $O_2$ by mixing nitrogen and dry air from two bottles. In the following, the concentration of oxygen will be given in $\%$ of the oxygen concentration of dry air and indicated with $\%$ air. This means, for example, that 20 $\%$ air was obtained by mixing 20 $\%$ dry air with 80 $\%$ nitrogen and therefore corresponds to 4 $\%$ vol $O_2$. The absolute error on the oxygen concentration adjusted with the gas mixing device is estimated to be below 1 $\%$ air. 
 
The excitation light was provided by a 405 nm LED (VAOL-5EUV0T4, VCC Visual Communications Company LLC), filtered by a shortpass (SP) filter with cut-off at 498 nm (498 SP BrightLine HC Shortpass Filter, Semrock) and focused on the surface of the samples with a collimation lens. The luminescence was focused by a lens and collected by a photodiode (SFH 213, Osram).
To suppress stray light and light reflected by the sample surface, the emission channel was equipped with a longpass filter with cut-off at 594 nm (594 LP Edge Basic Longpass Filter, Semrock) and a shortpass filter with cut-off at 682 nm (682 SP BrightLine HC Shortpass Filter, Semrock). The driver for the LED and the trans-impedance amplifier (TIA) are self-made.
For the frequency generation and the phase detection a two-phase lock-in amplifier (SR830, Stanford Research Inc.) was used. 

\subsubsection{Automated Data Acquisition}
\label{Data}

\begin{figure}[b!]
\centering
\includegraphics[keepaspectratio, width=5.8 cm]{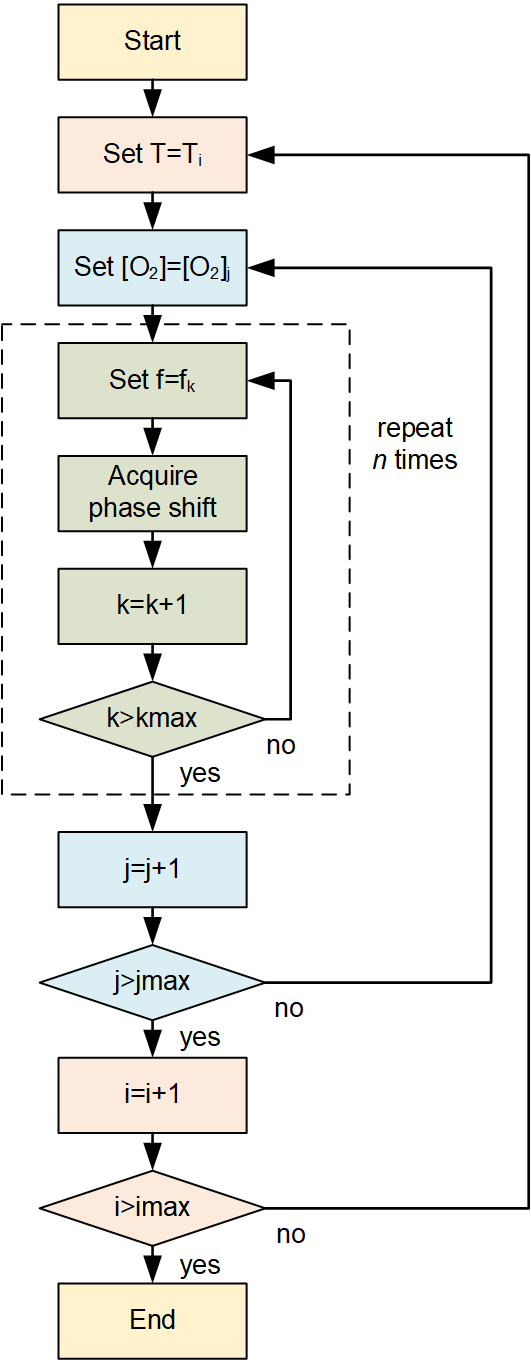}
\caption{Flow-chart of the automated data acquisition program.}
\label{fig:auto-data}
\end{figure}

The large amount of data needed for the training and the test of the neural network was acquired using an automated acquisition program which followed the flow-chart shown in Fig. \ref{fig:auto-data}. First, the program fixed the temperature and concentration. Then, the phase shift was measured for 50 modulation frequencies between 200 Hz and 15 kHz. This measurement was repeated 20 times. Next, keeping the temperature fixed, the program changed the oxygen concentration and the entire frequency-loop was repeated.
The oxygen concentration was varied between 0 $\%$ air and 100 $\%$ air in 5 $\%$ air steps.
Finally, the temperature was changed, and then the oxygen and frequency loops where repeated. The temperature was varied between 5 $^\circ$C and 45~$^\circ$C in 5 $^\circ$C steps.
The total number of measurements was thus 50 (frequencies) x 20 (loops) x 21 (oxygen concentrations) x 9 (temperatures) = 189'000, which required a total acquisition time of approximately 65 hours. This number of measurements was chosen as a compromise between maximizing the number of data and avoiding photodegradation, which naturally occurs when the sample is subjected to illumination. At the end of the session, a minimal change in the phase shift was observed.

\subsection{Neural Network Approach}
\label{NN}

The software component of this new sensor type is based on a neural network model (NNM). A NNM is made of three components \cite{Michelucci2017}: a neural network architecture (that includes how neurons are connected, the activation functions and all the hyperparameters), a loss function (here indicated with $L$) and an optimizer algorithm. In this section, those three components are described in detail.

\subsubsection{Neural Network Architecture}

\begin{figure}[t!]
\centering
\includegraphics[width=8.7 cm]{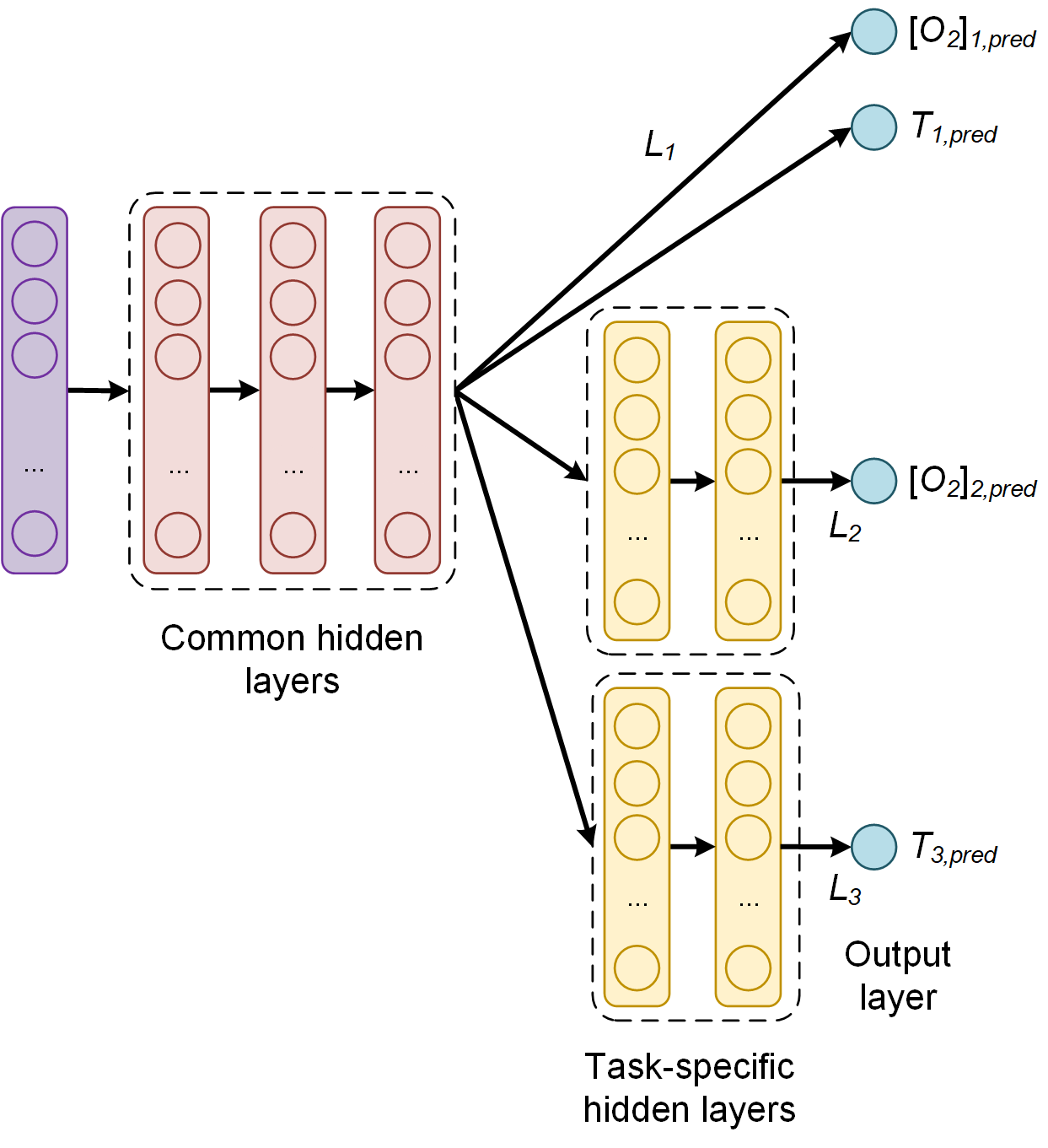}
\caption{Architecture of the multi-task learning neural network used in this paper. The common hidden layers generate a "shared representation" as output, that is used as input to task specific branches that learn specific features to each quantity and therefore improve the prediction accuracy. $L_i$ are the task-specific loss functions; $[O_2]_{i,pred}$ and $T_{i,pred}$ are the oxygen concentration and temperature predictions of  the corresponding branch $i$. Note that branch 2 and 3 have only one output.} 
\label{fig:NN_MTL_O2_T}
\end{figure}

The neural network used in this work has a multi-task-learning architecture and is depicted schematically in Fig. \ref{fig:NN_MTL_O2_T}. It consists of three {\sl common hidden layers} with 50 neurons each, which generates as output a "shared representation". The name shared representation comes from the fact that the output of common hidden layers is used to predict both $[O_2]$ and $T$. These layers are followed by three branches, one without additional layers to predict $[O_2]$ and $T$ at the same time, and two with each two additional {\sl task-specific hidden layers} to predict respectively $[O_2]$ and $T$. The shared representation is the input of two "task-specific hidden layers", that learn how to predict $[O_2]$ and $T$ better. This architecture uses the common hidden layers to find common features beneficial to each of the two tasks. During the training phase, learning to predict $[O_2]$ will influence the common hidden layers and, therefore, the prediction of $T$, and vice-versa. The further task-specific hidden layers learn features specific to each output and therefore improve the prediction accuracy. The number of neurons of each task-specific hidden layer used in this work is five. The activation function is the sigmoid function for all the neurons.  A study of which network architecture works best with this kind of data can be found in \cite{Michelucci2019_2}.

The network was trained with two types of input to test its effectiveness. In the first case, each observation consists of a vector of 50 values defined as
\begin{equation}
\label{input1}
{\pmb \theta}_s = \left(
\frac{\theta(w_1)}{90} , \frac{\theta(w_2)}{90} , ..., \frac{\theta(w_{50})}{90} 
\right)
\end{equation}
where $w_i$ are the 50 values of the angular modulation frequency of the excitation light (see Sec. \ref{Experimental}). The measured phase shift were divided by 90 to normalize the inputs between 0 and 1. In the second case, each observation is
\begin{equation}
\label{input2}
{\pmb \theta}_n = \left(
\frac{\theta(w_1)}{\theta_0(w_1)} , \frac{\theta(w_2)}{\theta_0(w_2)} , ..., \frac{\theta(w_{50})}{\theta_0(w_{50})} 
\right)
\end{equation}
where $\theta_0(w_i)$ is the value of the measured phase shift without oxygen quenching at the angular modulation frequency $w_i$.

\subsubsection{Loss Function}

The task-specific loss functions for each branch $i$ are indicated with $L_i$ and is the mean square error (MSE) defined as
\begin{equation}
L_i = \frac{1}{n} \sum_{j=1}^n \sum_{k=1}^{d_i} (y_{k,i}^{[j]}-\hat y_{k,i}^{[j]})^2, \ \ \ i=1,2,3
\label{MSE}
\end{equation}
where $n$ is the number of observations in the input dataset; ${\pmb y}_i^{[j]} \in \mathbb{R}^{d_i}$ is the measured value of the desired quantity for the $j^{th}$ observation, with $j=1, ..., n$  and $d_i$ is the dimension of the neural network branch output. In this case, $d_1=2, d_2=1$ and $d_3=1$. $ \hat {\pmb y}_i^{[j]} \in \mathbb{R}^{d_i}$ is the output of the network branch $i$, when evaluated on the $j^{th}$ observation. Since there are multiple branches, a global loss function $L$ is defined as a linear combination of the task-specific loss functions with weights $\alpha_i$ 
\begin{equation}
L = \sum_{i=1}^{n_T}\alpha_i L_i .
\label{globalcf}
\end{equation}
The parameters $\alpha_i$ have to be determined during the hyper-parameter tuning phase to optimize the network predictions.
In this paper, being the loss function the MSE (Eq. (\ref{MSE})), the global loss function is
\begin{equation}
L = \sum_{i=1}^{3}\alpha_i \frac{1}{n} \sum_{j=1}^n \sum_{k=1}^{d_i} (y_{i,k}^{[j]}-\hat y_{i,k}^{[j]})^2
\label{global_MSE}
\end{equation}
The global loss function weights used for this work were $\alpha_1 = 0.3$, $\alpha_2 = 5$ and $\alpha_3 = 1$. These parameters are the result of a hyper-parameter tuning for this architecture \cite{Michelucci2019_2}.

\subsubsection{Optimiser Algorithm}
\label{training}

The loss function was minimized using the optimizer Adaptive Moment Estimation (Adam) \cite{Kingma2014, Michelucci2017}. The implementation was performed using the TensorFlow\texttrademark $\ $library. The training was performed with a starting learning rate of $10^{-3}$. Two types of training were investigated to compare the training efficiency and performance of the network. {\sl No-batch training}: with this method all the training data  are used to perform an update of the weights and to evaluate the loss function. The loss function used is given by Eq. (\ref{global_MSE}). {\sl Mini-batch training}: with this method the weights update is performed after the network has seen 32 observations. In this case, Eq. (\ref{global_MSE}) is used with $n=32$. For each update of the weights, 32 random observations are chosen from the training dataset without repetitions until all the training data are fed to the network. 
The size of the mini-batch was chosen as a compromise between a good performance (small value of the loss function ) and the duration of training.

No-batch training has the advantage of stability and requires less time for each epoch since it performs one update of the weights using the entire training dataset. Mini-batch training is normally more effective in reaching small values of the loss function in less epochs, but it requires more time for each epoch \cite{Michelucci2017}. In our experiments for $20 \cdot 10^3$ epochs no-batch training took roughly five minutes on a modern MacBook Pro, while mini-batch training with $b=32$ took approximately 1 hour, thus resulting ca. 12 times slower.

\subsection{Performance Evaluation}

To evaluate the performance of the sensor, different metrics were analyzed. These are discussed in the next sections. The dataset $S$ of measured data was divided in two parts: one containing 80\% of randomly chosen observations (indicated with $S_{train}$), and one containing the remaining 20\% of the data (indicated with $S_{test}$). All the results presented were obtained by measuring the different metrics on the $S_{test}$ dataset.

\subsubsection{Absolute Error on the Prediction}

The metric used to compare predictions from expected values is the absolute error ($AE$) defined as the absolute value of the difference between the predicted and the expected value for a given observation. Note that in the architecture described in the previous sections, only branch 1 and 2 can predict $[O_2]$, 
while only branch 1 and 3 can predict $T$. The $AE$ for the oxygen concentration for the $j^{th}$ observation $[O_2]^{[j]}$  is 
\begin{equation}
\label{AE}
AE^{{[j]}}_{[O_2]} = |[O_2]^{{[j]}}_{pred}-[O_2]^{[j]}_{meas}|.
\end{equation}
where $[O_2]^{{[j]}}_{pred}$ and $[O_2]^{{[j]}}_{meas}$ are respectively the $[O_2]$ network prediction and measured value.
The further quantity used to analyse the performance of the network is the mean absolute error ($MAE$), defined as the average of the $AE$. For example, for the oxygen prediction using the training dataset $S_{train}$, the $MAE_{[O_2]}$ is defined as 
\begin{equation}
\label{MAE}
MAE_{[O_2]}(S_{train}) = \frac{1}{|S_{train}|} \sum_{j \in S_{train}}|[O_2]_{pred}^{[j]}-[O_2]_{real}^{[j]}|
\end{equation}
where $|S_{train}|$ is the size (or cardinality) of the training dataset. 
$AE_{T}$ and $MAE_T$ are similarly defined, using the prediction and the measured temperature values.

\subsubsection{Kernel Density Estimation}

A fundamental quantity to study the performance of the network is the prediction distribution of the $AE$s. This metrics carries information on the probability of the network to predict the expected value. To better illustrate this distribution, the kernel density estimate ($KDE$) of the distributions of the $AE$s was also calculated for both the oxygen concentration and the temperature. $KDE$ is a non-parametric algorithm to estimate the probability density function of a random variable by inferring the population distribution based on a finite data sample \cite{Hastie2009}.  In this work a Gaussian Kernel and a Scott bandwidth adaptive estimation \cite{Sain1996} using the seaborn Python package \cite{Waskom2020} were used.

\subsubsection{Error Limited Accuracy $\eta$}
\label{sektion:ela}

Generally, in a commercial sensor, the accuracy quantifies the performance of the sensor and helps to decide if the chosen device is appropriate for the application of interest. The above-defined metrics ($AE$, $MAE$ and $KDE$) are useful to compare the performance of different NNMs but do not help quantify which error the neural network senor will ultimately have in practice.
For this reason, in this work we introduce a new metric, called Error Limited Accuracy ($ELA$) and indicated with $\eta$.

\begin{definition*}
In a regression problem, given the metric $AE$, and a chosen value of it $\hat{AE}$, the $ELA$  $\eta$ limited by the error $\hat{AE}$ is defined as the number of predictions $\hat y$ of the NNM that lie in the range $|\hat y-y|\leq \hat{AE}$, with $y$ the expected value, divided by the total number of observations. It will be indicated with $\eta(\hat{AE})$. In more mathematical terms, given the set
\begin{equation}
E(\hat{AE}) = \{ \hat y^{[i]} \ {\text with } \ i = 1,..., n\ | \ \ |\hat y^{[i]}-y^{[i]}|\leq \hat{AE} \} 
\end{equation}
$\eta(\hat{AE})$ is defined as
\begin{equation}
\eta(\hat{AE}) = \frac{|E(\hat{AE})|}{n}
\end{equation}
where $|E(\hat{AE})|$ is the cardinality of the set $E(\hat{AE})$ or in other words, the number of its elements.
\end{definition*}

This metric allows interpreting the regression problem as a classification one. $\eta(\hat{AE})$ simply describes how many observations are predicted by the NNM within a given value of the absolute error. In other words, it represents the percentage of predictions that are within a certain error $\hat{AE}$ from the expected values. Finally, if we take $\hat{AE}$ big enough, all the predictions will be classified perfectly, so $\eta(\hat{AE})$ is expected to approach 1. The smaller $\hat{AE}$ is, the smaller will be the number of predictions correctly classified. We finally define $\overline{AE}$ as the value for which $\eta(\overline{AE})=1$, so the value of the absolute error for which the network predicts all the observations correctly. This value ($\overline{AE}$) will give us the biggest error in the sensor predictions.

\section{Results and Discussion}
\label{Results}

\subsection{Luminescence Experimental Results}

\begin{figure}[b!]
\centering
\includegraphics[width=8.2 cm]{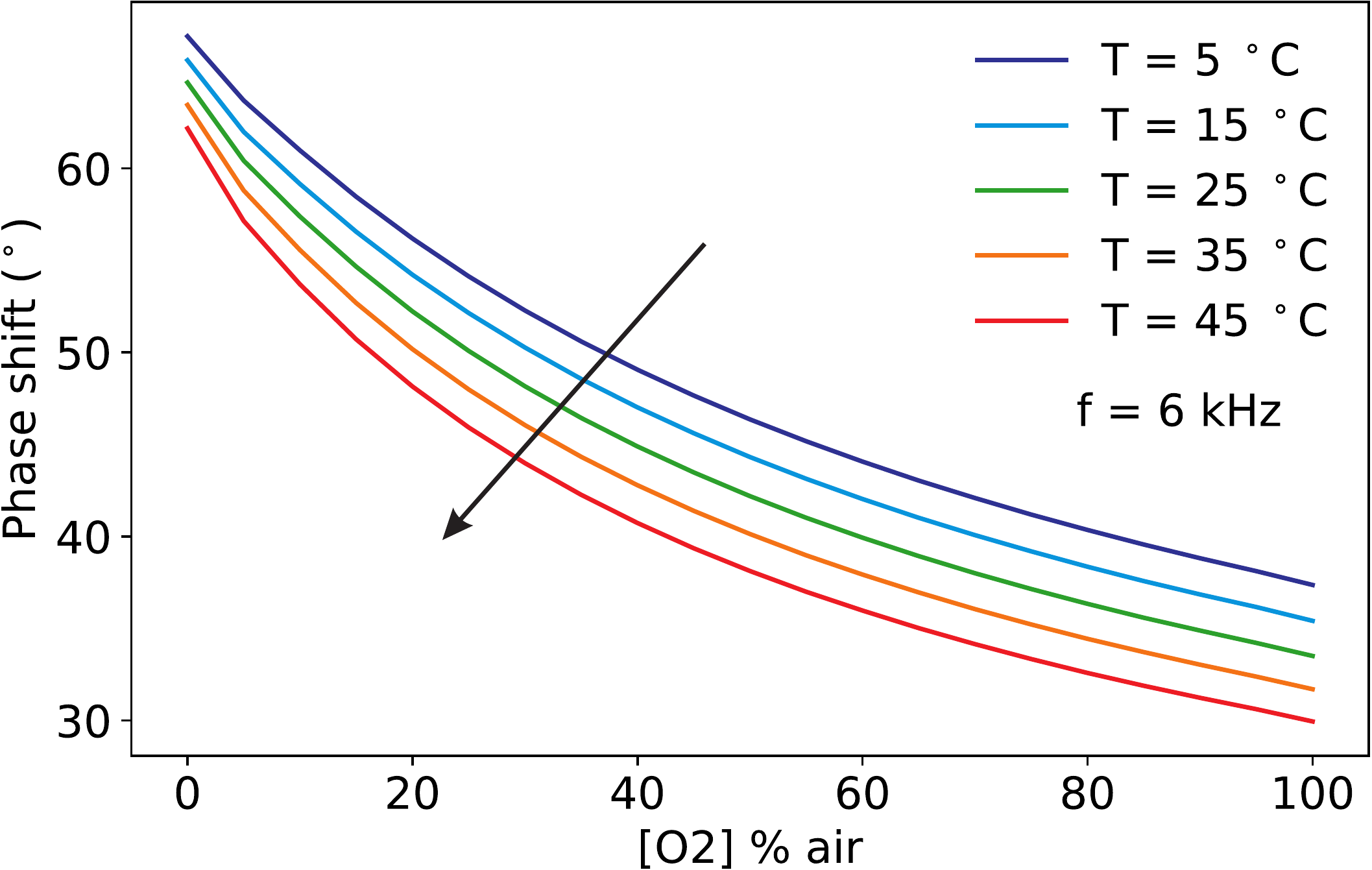}
\caption{Measured phase shift as a function of the oxygen concentration for selected temperatures at a fixed modulation frequency of 6 kHz. The arrow marks increasing temperatures.}
\label{fig:expdata1}
\end{figure}

\begin{figure}[t!]
\centering
\includegraphics[width=8.2 cm]{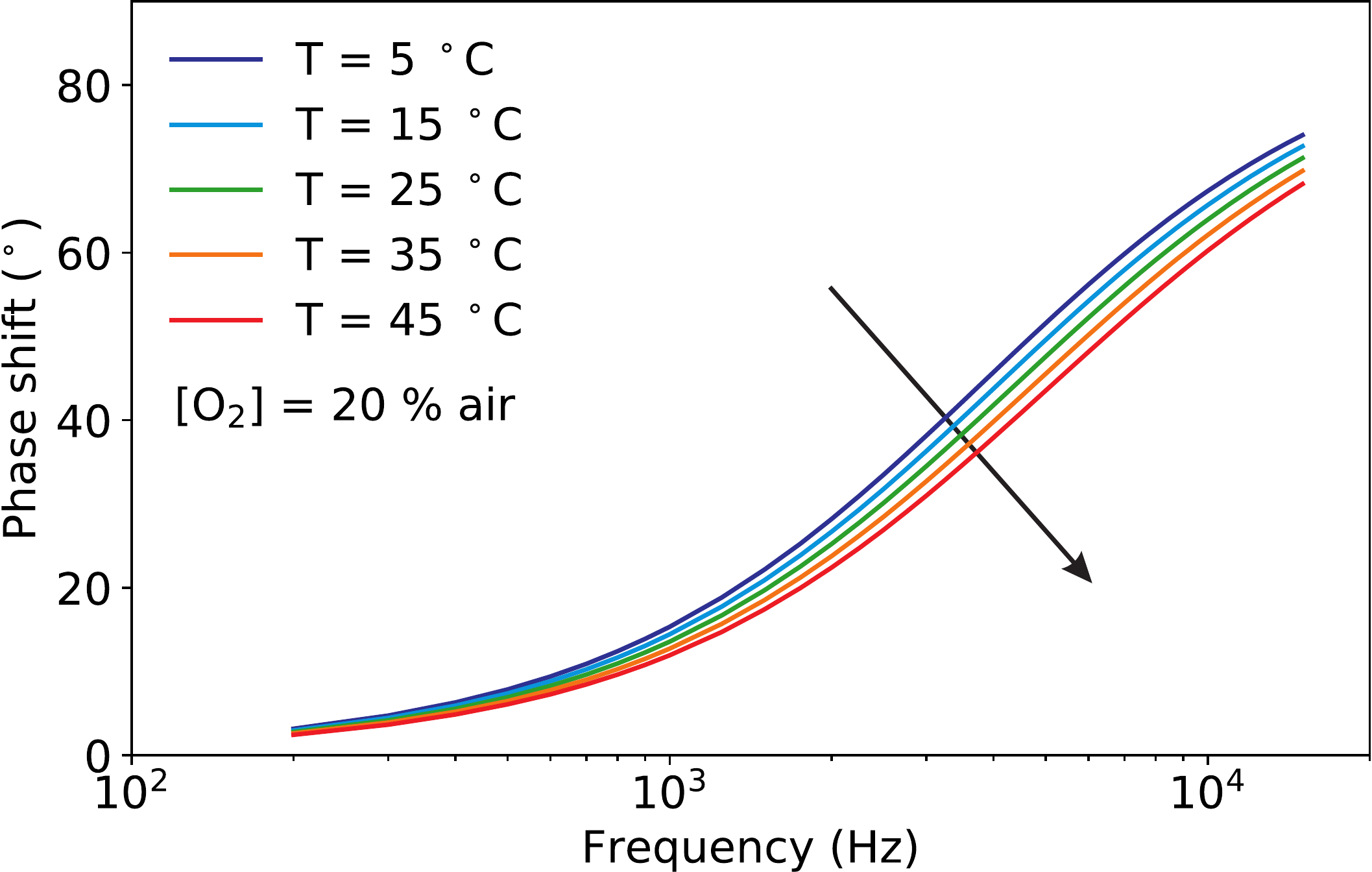}
\caption{Measured phase shift as a function of the modulation frequency for selected temperatures at a fixed oxygen concentration of $[O_2]=20 \  \%$ air. The arrow marks increasing temperatures.}
\label{fig:expdata2}
\end{figure}

\begin{figure}[t!]
\centering
\includegraphics[width=8.2 cm]{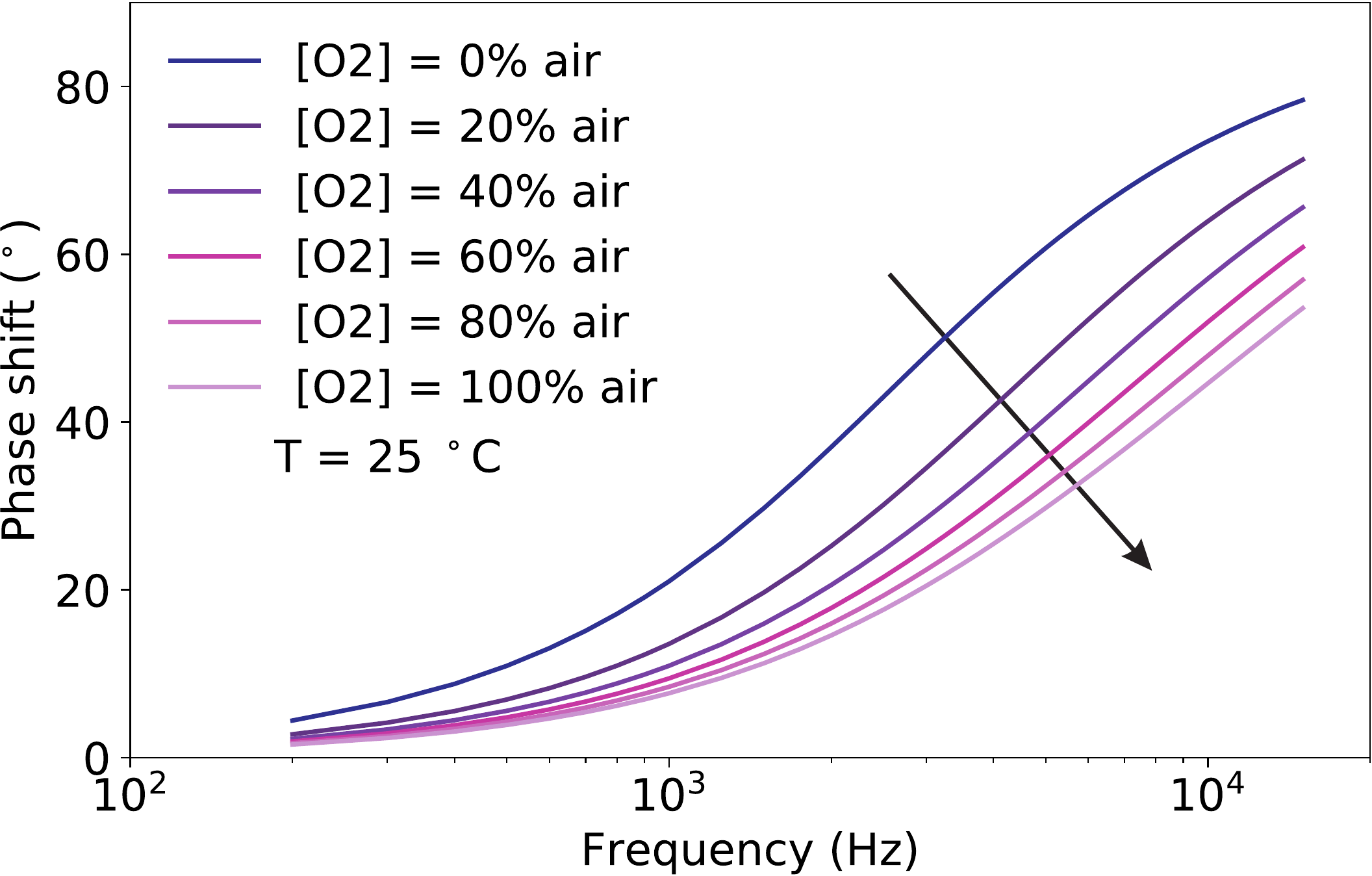}
\caption{Measured phase shift as a function of the modulation frequency for selected oxygen concentrations at a fixed temperature of $T=25 \ ^{\circ}$C. The arrow marks increasing oxygen concentrations.}
\label{fig:expdata3}
\end{figure}

As described in Section \ref{Theory}, the phase shift depends non-linearly on the oxygen concentration according to the Stern-Volmer equation. It depends also on the temperature, which influences the luminescence and the collision mechanisms, and on the modulation frequency of the excitation light, as described in Eq. (\ref{theta_full}). The experimental observations for the phase shift for variations of these three quantities are shown in the Figs. \ref{fig:expdata1} to \ref{fig:expdata3}.

Fig. \ref{fig:expdata1} shows the measured phase shifts as a function of the oxygen concentration at a constant modulation frequency of 6 kHz and for increasing temperatures. For clarity, the results at only few selected temperatures are shown. The decrease of the phase shift due to the collisional quenching is clearly visible in all curves. The phase shift is, as expected, also strongly  temperature-dependent. For $[O_2]=0$, in the absence of oxygen, the reduction of the phase shift with increasing $T$ is due to temperature quenching; the influence of temperature becomes stronger at higher oxygen concentration, as a result of the increase of the diffusion rates of oxygen through the sample.

For a given oxygen concentration, the phase shift is strongly dependent on the modulation frequency, as it can be seen in Fig. \ref{fig:expdata2}, where the shape of the frequency response is determined by the distribution of decay times of the sample. From the figure it is visible that the reduction of the phase shift with increasing temperatures is not constant but depends on the modulation frequency.

For completeness, the effect of the oxygen concentration on the frequency response at a fixed temperature is shown in Fig. \ref{fig:expdata3}. Compared to Fig. \ref{fig:expdata2}, the frequency response of the sample is affected more strongly by the oxygen concentration than by temperature. In other words, the sample has a higher sensitivity to oxygen than to temperature.

The measurements of Figs. \ref{fig:expdata1} to \ref{fig:expdata3} show how similar the curves of the phase shift are for different values of oxygen, temperature and modulation frequency. This helps to understand why it is not possible from the measurement of the phase shift, or even of the phase shift for varying modulation frequencies, to simultaneously determine both the oxygen concentration and the temperature using Eq. (\ref{theta_full}). The temperature must be known in advance and used to compute the oxygen concentration. This is no longer the case with the neural network approach, as it will be shown in the next section.

\subsection{Sensor Performance}

\begin{figure*}[htbp]
\centering
\includegraphics[width=15 cm]{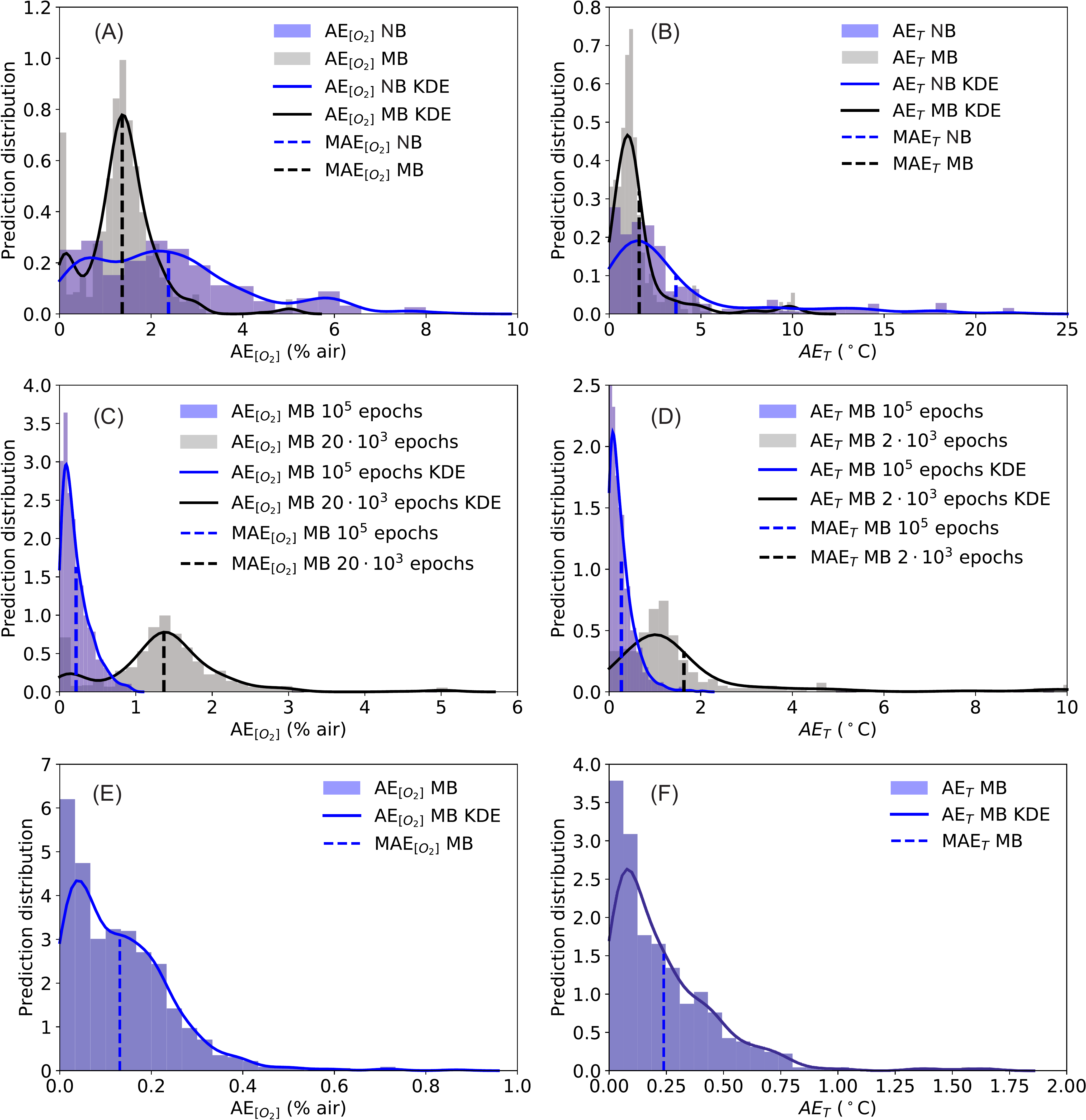}
\caption{Distributions of the neural network predictions for the oxygen concentration (panels (A), (C) and (E)) and for the temperature (panels (B), (D) and (F)). In all panels the normalized prediction distribution histogram (columns), the kernel density estimate ($KDE$) of the distribution of the $AE$s (solid line), and $MAE$ (dashed vertical line) are shown. Panels (A) and (B): Comparison between training using no batches (NB) and using mini-batches (MB) with a batch size of 32 for 20'000 epochs; the input of the network is ${\pmb \theta}_s$. Panels (C) and (D): Comparison between training using mini-batches (MB) with a batch size of 32 for 100'000 and 20'000 epochs; the input of the network is ${\pmb \theta}_s$. Panels (E) and (F): 
training using mini-batches (MB) with a batch size of 32 for 20'000 epochs; the input of the network is ${\pmb \theta}_n$.}
\label{fig:KDE_results_all}
\end{figure*}

First, the effect of the training on the sensor performance was investigated. As described in Section \ref{training}, the neural network was trained with no-batches and with mini-batches. For this comparison the network was trained for 20'000 epochs using the input observations ${\pmb \theta}_s$ as defined in Eq. (\ref{input1}). The results for $AE_{[O_2]}$ and $AE_T$ are shown in Fig. \ref{fig:KDE_results_all}(A) and \ref{fig:KDE_results_all}(B), respectively. The blue histogram shows the $AE$ distribution when using no-batch, the gray when using mini-batches of size 32. The $KDE$ profiles help illustrating the features of the histogram. The effect of introducing mini-batches on the performance is significant. The predictions distributions get much narrower, the mean average errors decrease from $MAE_{[O_2]}=2.4$ \% air and $MAE_{T}=3.6 \ ^\circ$C to $MAE_{[O_2]}=1.4$ \% air and $MAE_{T}=1.6 \ ^\circ$C. Although the performance is significantly improved, from Fig. \ref{fig:KDE_results_all}(A) and \ref{fig:KDE_results_all}(B) it can also be clearly seen that errors as high as approximately 5~\%~air for $[O_2]$ or 12 $^\circ$C for $T$ are possible.

\begin{figure*}[t!]
\centering
\includegraphics[width=14 cm]{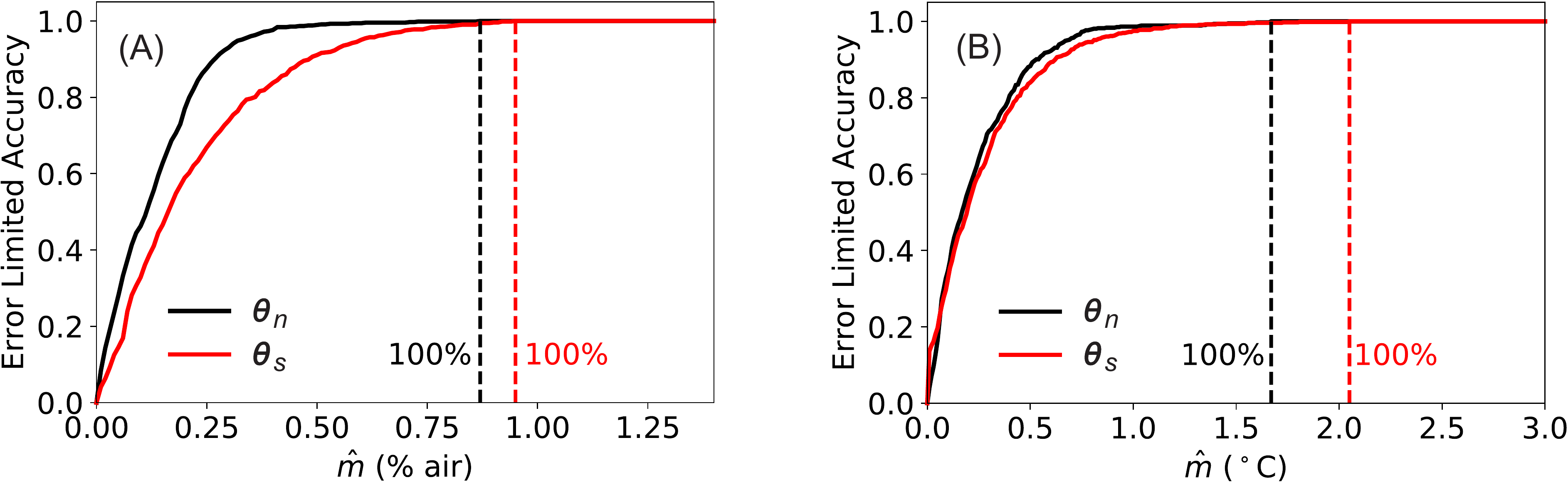}
\caption{Comparison of the $ELA$ $\eta$: Panel (A) oxygen prediction, panel (B) temperature prediction. The black lines are the results obtained with a network that was trained with ${\pmb \theta}_n$ as input for 20'000 epochs with mini-batchs of size 32, while the red ones with ${\pmb \theta}_s$ as input for 100'000 epochs with mini-batchs of size 32. The dashed lines indicates the values of the $\overline{AE}$ for which the predictions would give $\eta=1$.}
\label{fig:ELA_result_comparison}
\end{figure*}

Fig. \ref{fig:KDE_results_all}(C) and \ref{fig:KDE_results_all}(D) show the comparison between prediction distributions with 20'000 and 100'000 epochs (always using a mini-batch of size 32), using the input observations ${\pmb \theta}_s$ as defined in Eq. (\ref{input1}). The effect of longer training is a dramatic improvement in the performance. When the network was trained for 100'000 epochs the mean average errors are reduced to only $MAE_{[O_2]}=0.22$ \% air and $MAE_{T}=0.27  \ ^\circ$C. Additionally, all the predictions for $[O_2]$ lie below 0.94 \% air, and for $T$ lie below 2.1 $^\circ$C.

The results of Fig. \ref{fig:KDE_results_all}(C) and \ref{fig:KDE_results_all}(D) demonstrate two new findings: 1) with the proposed approach, it is possible to predict both $[O_2]$ and $T$ at the same time from the phase shift using a single luminophore; 2) the prediction has an expected error which is comparable or below the typical accuracy of commercial sensors. The possibility of dual sensing paves the road to the development of a completely new generation of sensors.
The price to pay is that the training of a network for 100'000 epochs requires approximately 5 hours on a modern laptop.

To investigate if the training can be performed more efficiently, the normalized phase shift ${\pmb \theta}_n$ defined in Eq. (\ref{input2}) is used as input to the network. The performance of the network in this case, with a mini-batch size of 32 and a training of 20'000 epochs is shown in Fig. \ref{fig:KDE_results_all}(E) and \ref{fig:KDE_results_all}(F). The performance is further improved: even if the number of epochs is only 20'000 the mean average errors are better than what obtained with ${\pmb \theta}_s$ and a training of 100'000 epochs, achieving $MAE_{[O_2]}=0.13$ \% air and $MAE_{T}=0.24 \ ^\circ$C. The distributions are also narrower, particularly for the temperature. Additionally, all the $AE_{[O_2]}$ lie below 0.87 \% air, and  $AE_{T}$ below 1.7 $^\circ$C. This type of training is clearly more efficient. The reason may lie in the additional information which is fed to the network when using the input ${\pmb \theta}_n$ and in the simplified functional behavior of ${\pmb \theta}_n$ compared to ${\pmb \theta}_s$ as it may be expected by Eq. (\ref{theta_full}).

The performance of the different neural networks is summarized in Table \ref{TableMAE_summary}. 
\begin{table}[hbt]
\centering
\caption {\bf Summary of the performance for neural network models}

\begin{tabular}{ cccc}
\smallskip 
 Input & Epochs / Batch size & $MAE_{[O_2]}$ & $MAE_{[T]}$  \\ 
 \hline
${\pmb \theta}_s$ & 20'000 / \textrm{no batch} & 2.4 \% air & 3.6 $^\circ C$\\ 
${\pmb \theta}_s$ & 20'000 / 32 & 1.4\% air & 1.6 $^\circ C$\\ 
${\pmb \theta}_s$& 100'000 / 32 & 0.22 \% air & 0.27 $^\circ C$\\ 
${\pmb \theta}_n$ & 20'000 / 32 & 0.13 \% air & 0.24 $^\circ C$\\ 
\end{tabular}
\label{TableMAE_summary}
\end{table}

\subsection{Error Limited Accuracy}

The metrics discussed in the previous sections are useful to compare the network performance and to measure how good the predictions are. However, they do not offer an understanding on what a sensor built with such a model could achieve. For practical applications, the relevant question is rather what is the maximum error which the sensor will have predicting the oxygen concentration and temperature. To answer this question, the $ELA$ ($\eta$) defined in Section \ref{sektion:ela} can be used. As explained previously, $\eta$ is defined depending on the chosen metric $m$. In this section, the metric chosen is $m=AE_{[O_2]}$ for the oxygen concentration and $m=AE_{T}$ for the temperature. This new metrics will allow the determination of the maximum error of the sensor. 

Fig. \ref{fig:ELA_result_comparison} displays the $ELA$ $\eta(\widehat {AE})$ for oxygen concentration (A) and for the temperature (B). In each panel, the results obtained using the input ${\pmb \theta}_n$ and a training for 20'000 epochs are shown in black, and the results obtained using the input ${\pmb \theta}_s$ and a training for 100'000 epochs in red. In both cases, the training was performed with mini-batches of size 32.
The dashed lines indicate the values of the $\overline{AE}_{[O_2]}$ and $\overline{AE}_{T}$ for which the error limited accuracy $\eta$ equals 1. In other words, all the predictions will have an error equal or smaller than $\overline{AE}$.

\begin{table}[t!]
\centering
\caption {\bf Summary of the values of $\overline{AE}$ for the cases shown in Fig. \ref{fig:ELA_result_comparison}(A) and \ref{fig:ELA_result_comparison}(B).}
\begin{tabular}{ cccc}
\smallskip 
 Input & Epochs / Batch size & $\overline{AE}_{[O_2]}$ & $\overline{AE}_{T}$  \\ 
 \hline
${\pmb \theta}_s$ & 100'000 / 32 & 0.95 \% air & 2.1 $^\circ C$\\ 
${\pmb \theta}_n $ & 20'000 / 32 & 0.87\% air & 1.7 $^\circ C$\\ 
\end{tabular}
\label{table:ela}
\end{table}

Fig. \ref{fig:ELA_result_comparison}(A) shows that, for the network trained with ${\pmb \theta}_s$ as input, the model would predict perfectly all the oxygen concentrations within 0.95 \% air error. For the network trained with ${\pmb \theta}_n$ this value is futher reduced to 0.87 \% air. $\overline{AE}_{[O_2]}$ can be interpreted as the accuracy a sensor based on this NNM  would have.
Fig. \ref{fig:ELA_result_comparison}(B) shows the results of the same analysis for the temperature measurement. The interpretation is similar to the one given above for the oxygen concentration. For the network trained with ${\pmb \theta}_s$ as input, the model would predict perfectly all the temperature values within $\overline{AE}_{T}=2.1 \ ^\circ$C error. For the network trained with ${\pmb \theta}_n$ this value would be  $\overline{AE}_{T}=1.7 \ ^\circ$C. The values of $\overline{AE}_{[O_2]}$ and $\overline{AE}_{T}$ are summarized in Table \ref{table:ela}.

\section{Conclusions}

In this work, a new sensor learning approach to luminescence sensing is presented. The proposed method allows parallel inference, or the extraction of multiple physical quantities simultaneously, from a single dataset without any {\sl a priori} mathematical model, even in the presence of cross interferences. Classical approaches to this type of problems in physics can be challenging or impossible to solve if the mathematical models describing the functional dependencies are too complex or even unknown.

The approach is demonstrated by realizing a luminescence sensor, which uses a single luminophore and a single measuring channel, and can measure simultaneously both the oxygen concentration and the temperature of a medium. This is achieved using a multi-task learning neural network model, which was trained on a very large dataset. The results in the prediction of the oxygen concentration and temperature show unprecedented accuracy for both parameters, demonstrating that this approach can make a new generation of dual- or even multiple-parameter sensors possible.
The expected error or accuracy of a sensor based on a given NNM approach is intrinsically difficult. For this reason, the new metric Error Limited Accuracy $ELA$ ($\eta(AE)$) is proposed. The $ELA$ enables to estimate how many predicted values lie within a certain absolute error from the expected measurement. This new metric allows therefore giving a maximum measurement error of the NNM results.

The ability to predict both $[O_2]$ and $T$ at the same time, from a single set of data obtained with a single indicator, has profound implications for the development of luminescence sensors. Sensors will become easier and cheaper to build since no separate temperature measurements are necessary anymore. Generally, the effect of interferences can be learned by the neural network and do not need to be corrected for in the data processing. 

This work opens the road to complete new optical sensing approaches for future generations of sensors. Those sensors will be able to extract multiple physical quantities from a common set of data at the same time to achieve consistent results that are both accurate and stable. The described approach is relevant for many practical applications in sensor science and demonstrates that this model-free approach has the potential of revolutionizing optical sensing.

%\section*{Funding Information}

%\section*{Acknowledgments}

%\section*{Disclosures}

\medskip

\noindent\textbf{Disclosures.} The authors declare no conflicts of interest.

% Bibliography
\bibliographystyle{elsarticle-num-names}
\bibliography{bibliography}

% Full bibliography will be added automatically on a new page for Optics Letters submissions. This command is ignored for journal article submissions.
% Note that this extra page will not count against page length.
%\bibliographyfullrefs{bibliography}

%\printbibliography

%Manual citation list
%\begin{thebibliography}{1}
%\bibitem{Michelucci2017}
%Michelucci, U.
%{\sl Applied Deep Learning - A Case-Based Approach to Understanding Deep Neural Networks}; Apress Media, LLC: New York, NY, USA, 2018; pp. 374--375.

%\bibitem{Kingma2014}
%Kingma, D.P.; Ba, J.
%Adam: A method for stochastic optimization. In Proceedings of 3rd International Conference on Learning Representations, ICLR 2015, San Diego, CA, USA, May 7-9, 2015, pp. 1--15.

%\bibitem{Michelucci2019}
%Michelucci, U.; Baumgartner, M.; Venturini, F.
%Optical oxygen sensing with artificial intelligence.
%{\sl Sensors} {\bf 2019}, {\sl 19}, 777.

˜
%\bibitem{Zhang:14}
%Y.~Zhang, S.~Qiao, L.~Sun, Q.~W. Shi, W.~Huang, %L.~Li, and Z.~Yang,
 % \enquote{Photoinduced active terahertz metamaterials with nanostructured
  %vanadium dioxide film deposited by sol-gel method,} Opt. Express \textbf{22},
  %11070--11078 (2014).
%\end{thebibliography}

\end{document}